\documentclass[reprint,onecolumn,longbibliography,floats,aps,amsmath,amssymb,nofootinbib,prd,notitlepage]{revtex4-2}

\usepackage{graphicx}
\usepackage{amsmath,amssymb}
\usepackage{hyperref}
\usepackage{graphicx}
\usepackage{subfigure}
\usepackage{arydshln}
\usepackage{inputenc}
\usepackage{graphicx}
\usepackage{amsmath,amssymb}
\usepackage{hyperref}
\usepackage{appendix}
\usepackage{enumerate}
\usepackage{color}

\begin{document}
\title{The Primodial Power Spectrum in Loop Quantum Cosmology for different regularizations}

\author{Maciej Kowalczyk}
\email{maciej.kowalczyk@uwr.edu.pl}
\affiliation{University of Wrocław, Faculty of Physics and Astronomy, Institute for Theoretical Physics, pl. M. Borna 9, 50-204 Wrocław, Poland}

\author{Guillermo A. Mena Marug\'an}
\email{mena@iem.cfmac.csic.es}
\affiliation{Instituto de Estructura de la Materia, IEM-CSIC, Serrano 121, 28006 Madrid, Spain}

\author{Tomasz Paw{\l}owski}
\email{tomasz.pawlowski@uwr.edu.pl}
\affiliation{University of Wrocław, Faculty of Physics and Astronomy, Institute for Theoretical Physics, pl. M. Borna 9, 50-204 Wrocław, Poland\,}

\begin{abstract}
In Loop Quantum Cosmology, the quantization of the Hamiltonian constraint involves a regularization procedure which is affected by certain ambiguities. Moreover, different regularizations lead to distinct mathematical formulations and, consequently, to different physical predictions. In this work, we explore the impact of this regularization on the primordial power spectrum of cosmological perturbations. More specifically, we study this power spectrum for the conventional regularization used in Loop Quantum Cosmology and for two alternative prescriptions suggested in the literature. We set initial conditions for the perturbations at the bounce corresponding to a recently proposed vacuum state, optimally adapted to the background dynamics. This choice of vacuum is based on an asymptotic Hamiltonian diagonalization in the ultraviolet sector of the perturbations which provides a non-oscillating power spectrum. Employing a suitable approximation to the propagation equations of the perturbations around the bounce, we are able to obtain an analytic expression for the primordial power spectrum of this vacuum for all the discussed regularizations. We compare the results and prove that the main relevant  distinction between the corresponding spectra is the scale where power suppression occurs. The associated wavenumber scale is proportional to the square root of the critical density in Loop Quantum Cosmology, density which is different for each of the studied regularizations.
\end{abstract}

\maketitle
\section{Introduction}

The quest to understand the origin and evolution of the Universe is one of the fundamental challenges of modern cosmology. Over the past decades, the inflationary paradigm \cite{Baumann:2009ds,Guth:1980zm, Linde:1981mu, Starobinsky:1979ty,Sato:1981ds} has become a foundational ingredient in this endeavor, giving a compelling explanation for the homogeneity observed in the Universe. It also provides a mechanism for generating the primordial fluctuations \cite{Mukhanov:2005sc} that seed cosmic microwave background (CMB) anisotropies and large-scale structure formation. The standard framework is provided by chaotic inflation, in which the initial energy density of the inflaton is of Planck order \cite{Linde:1983gd}. High-precision observations \cite{Planck:2013pxb, Planck:2018jri, Planck:2018vyg, WMAP:2012nax} of the CMB support the predictions of inflation, particularly the concept of a slow-roll phase, where the inflaton field driving the expansion evolves slowly with a nearly constant potential. However, according to these observational data, the energy density associated with this slow-roll phase is approximately twelve orders of magnitude smaller than the Planck energy density \cite{Liu:2024zql}, a discrepancy that underscores the tension between the simplest inflationary models and the extreme conditions of the early Universe. This mismatch raises questions about the compatibility of the standard framework with a genuine quantum origin for the vacuum fluctuations, and has even been claimed to lead to problems with homogeneity if the gravitational phenomena at the two aforementioned energy density scales are combined using models with delayed inflation \cite{Liu:2024zql}. 

An appealing approach to address these issues is to assume first the applicability of perturbation theory in cosmology and then incorporate the quantum nature of gravity in the field description of those perturbations. This gravitational quantum behavior is expected to play a pivotal role in shaping spacetime dynamics at the scales involved in the very early Universe. One of the more popular candidates for achieving a quantum description of this spacetime geometry is Loop Quantum Cosmology (LQC) \cite{agullo2023loop, Ashtekar:2011ni}. LQC adapts the quantization methods of Loop Quantum Gravity (LQG) \cite{Thiemann:2001gmi,LQG} to simplified, symmetry-reduced cosmological models. Apart from that, it provides a natural mechanism for resolving the Big Bang singularity by means of a quantum bounce \cite{Ashtekar:2006rx}, characterized by a specific (critical) value of the matter energy density, which is the maximum allowed in LQC for this density.

However, the quantization of the geometry carried out in LQC involves a regularization of the Hamiltonian, a procedure which has been argued to be affected by ambiguities. Different regularizations lead to models with differing mathematical structures and, in part, to distinct physical predictions. More specifically, restricting our attention to the context of the simplest isotropic models in cosmology, essentially three regularization prescriptions have been proposed and investigated. These differ in how the so-called Lorentzian term of the Hamiltonian constraint is expressed using holonomies and the spatial volume. The first approach \cite{Ashtekar:2006rx} decomposes this term into a linear combination of the spatial Ricci scalar and the curvature part of the Hamiltonian constraint. We will refer to the cosmological model built with this regularization as {\emph {conventional}} LQC, because it is the regularization usually adopted and explored in more detail. The second approach follows Thiemann's proposal in full LQG to re-express the exterior curvature in terms of the more elementary holonomy and triad variables \cite{Thiemann:2001gmi}. The homogeneous and isotropic cosmological quantum model obtained with this regularization is often called mLQC-I in the literature \cite{Yang:2009fp}. Finally, the third approach applies a first-order approximation of the curvature in terms of the Ashtekar connection. The associated cosmological model is often referred to as mLQC-II \cite{Yang:2009fp}. Although all these approaches converge to the predictions of classical general relativity in a semiclassical regime for large volumes, they may yield different outcomes in the high-energy regime. One such example is the value of the critical density at the bounce, which is different for each of the three cases. 

In LQC, this bounce is followed by a Planck epoch of superinflation, characterized by a rapid increase in the Hubble parameter to values approaching the Planck scale. This epoch is driven by quantum effects that modify the dynamics of the Universe, setting the stage for subsequent phases of evolution. Once the energy density drops well below its critical value, these quantum corrections diminish. In effective inflationary solutions of phenomenological interest in LQC \cite{Ivan,hybr-pred,AGvacio2}, this quantum epoch gives then way to a phase dominated by the kinetic energy of the inflaton, where the Universe undergoes a power-law expansion. In this period, the classical description of general relativity becomes valid. This kinetically dominated phase naturally transitions into a standard slow-roll inflationary phase, where the potential energy of the inflaton becomes the primary driver of expansion. Notably, this evolution of the background can be regarded to a certain extent as a model with delayed inflation.  We refer the reader to Ref. \cite{PhysRevD.110.103502} for a recent discussion of this consideration.

The study of perturbations in LQC offers additional insights into the connection between the effects of quantum physics deep in the Planck regime and cosmological observations. For instance, considerable attention has been devoted to understanding how quantum modifications, in the bounce regime, influence the evolution of perturbations and their observable imprints on the CMB angular power spectrum \cite{Ashtekar:2021izi,Li:2024xxz}. These investigations reveal that the quantum geometry effects during the bounce not only modify the dynamics of primordial scalar and tensor perturbations but can also potentially address key observational challenges, such as the suppression of power at large angular scales in the CMB. The modifications are encoded in the pre-inflationary dynamics, where the choice of initial conditions plays a crucial role. These initial conditions should be those corresponding to a vacuum state, adapted optimally to the background dynamics. Thus, as a near bounce behavior of the background will drastically differ from a de Sitter solution, it turns out that the vacuum which is always taken as a preferred state in a quasi-de Sitter inflationary epoch, namely the Bunch-Davies vacuum \cite{Bunch:1978yq}, ceases to be an optimal choice. To generalize this choice, a criterion based on fundamental physical principles has been recently proposed to determine a privileged vacuum in cosmological preinflationary and inflationary scenarios  \cite{ElizagaNavascues:2019itm}. It is based on an asymptotic Hamiltonian diagonalization (ADH, from the initials) for the perturbations in the ultraviolet sector which results in a non-oscillating (NO) Primordial Power Spectrum (PPS). Owing to these properties, the corresponding state is known in the literature as NO-AHD vacuum.

The conventional LQC model with a choice of NO-AHD vacuum has been investigated in detail in Refs. \cite{Navascues:2021mxq,ElizagaNavascues:2023xah,MenaMarugan:2024vyy,MenaMarugan:2024zcv} using a hybrid quantization scheme for the combined system of the background and the perturbations \cite{ElizagaNavascues:2020uyf} (a similar analysis for an alternative quantization scheme, called the dressed metric approach \cite{Agullo:2012sh,Agullo:2012fc,Agullo:2013ai}, can be found in Ref. \cite{Alonso-Serrano:2023xwr}). This study has been carried out analytically and then confirmed by numerical means, proving the goodness of the proposed analytic approximations. The goal of our work in this paper is to extend such an (approximated) analytic study of the PPS of the perturbations to the other two aforementioned regularizations of the Hamiltonian in LQC, leading respectively to the mLQC-I and mLQC-II models. In particular, we are interested in investigating how different regularizations affect the scale of power suppression in the resulting PPS and the shape of this spectrum.

The rest of the paper is organized as follows. We begin in Sec. \ref{section-I} by briefly summarizing the dynamics of the effective cosmological background in LQC for the three different regularizations. In Sec. \ref{cosmo} we introduce the background dependent mass of the perturbations in the hybrid quantization approach (deep in the Planck regime), using a P\"oschl-Teller approximation as suggested in Ref. \cite{Navascues:2021mxq}. We also present the NO-AHD vacumm state. Then, in Sec. \ref{section-PPS} we derive and plot the corresponding PPS for each of the three regularizations. Finally, we discuss our results in Sec. \ref{section-Discussion} and compare them among themselves and with the PPS obtained in Ref. \cite{PhysRevD.110.103502} for delayed inflation with initial and final de Sitter epochs connected by a kinetically dominated expansion. In the following, we use Planck units, setting $G$, $c$, and $\hbar$ equal to one.

\section{Effective cosmological background}\label{section-I}

In LQC, the gravitational sector is described using phase space variables inherited from LQG, corresponding to holonomies and densitized triads. The symmetry reduction to homogeneous and isotropic spacetime considerably simplifies the gravitational system, in particular leaving the Hamiltonian constraint as the only remaining nontrivial constraint. It adopts the form
 \begin{equation}
\mathcal{H}_{g}=\mathcal{H}^E-2(1+\gamma^2)T ,
\end{equation}
where $\mathcal{H}^E$ and $T$ respectively represent the so-called Euclidean and Lorentzian contributions, because the latter would be absent if the spacetime signature were positive. The Euclidean part is related to the intrinsic curvature, whereas the Lorentzian term incorporates extrinsic curvature contributions. On the other hand, $\gamma$ is the Immirzi parameter, which usually is fixed to the value $\gamma \approx 0.2375$ obtained from black-hole entropy considerations (see e.g. Ref. \cite{LQG}).

The quantization process in LQC is affected by some ambiguities, particularly concerning the treatment of the Lorentzian term in the Hamiltonian constraint. In conventional LQC, a notable simplification is employed: for flat cosmological models, the Lorentzian term is classically proportional to the Euclidean one. This proportionality allows the Lorentzian term to be reformulated in terms of the Euclidean part before quantization, streamlining the quantum Hamiltonian structure. As we mentioned above, alternative approaches to handling the Lorentzian term have been proposed in Ref. \cite{Yang:2009fp}. The first one, that leads to the mLQC-I cosmological model, is rooted in the algorithm introduced by Thiemann \cite{Thiemann:2001gmi} for full LQG, in which the extrinsic curvature is re-expressed using extended holonomies and fluxes. Such regularization has closer correspondence to the methodology of full LQG and offers a richer dynamical structure, but with increased complexity in the quantum description. The second alternative is based on a simple first order approximation of the curvature operator in terms of the Ashtekar connection. This regularization provides the mLQC-II model.

The mathematical structure and the quantum cosmological evolution have already been investigated in detail for all these regularizations \cite{Ashtekar:2006rx, Assanioussi:2019iye,Kowalczyk:2022ajp}. Here we summarize the results about the behavior of their associated effective solutions \cite{Li:2018fco, Li:2021mop} when backreaction is totally neglected. Assuming a flat isotropic Friedmann–Lemaître–Robertson–Walker (FLRW) spacetime with a homogeneous scalar field as the matter content, the Universe dynamics emerges from the vanishing of the total Hamiltonian constraint, given by the gravitational part $\mathcal{H}_g$ and the matter field contribution. In this way, one obtains modified evolution equations for the scale factor. In the case of conventional LQC, they read
\begin{equation}\label{eq:Friedmann-LQC}
    H^2=\frac{8\pi }{3}  \rho  \left(1-\frac{\rho }{\rho_c}\right), \qquad \frac{\ddot{a}}{a} = -\frac{4\pi }{3} \rho \left(1 - 4\frac{\rho}{\rho_c}\right) - 4\pi  P \left(1 - 2\frac{\rho}{\rho_c}\right),
\end{equation}
where $H$ is the Hubble parameter, $\rho_c=0.4092$ is the critical energy density at the bounce for this conventional LQC framework, and $\rho$ and $P$ are the energy density and pressure of the matter field content, respectively. Importantly, the matter-energy conservation law remains valid. When $\rho \ll \rho_c$, the above modified equations reduce to their classical counterpart, recovering the standard cosmological dynamics. Assuming that the main contribution to the energy density of the scalar field is a kinetic term, the evolution of the scale factor can be solved analytically in proper time, $t$, 
\begin{equation}\label{eq:scalefactor-lqc}
a(t)=(1 + 24 \pi \rho_c t^2 )^{1/6}.
\end{equation}
Here, and without loss of generality, we have chosen $t=0$ at the bounce for convenience and we have set the global scale of distances such that the scale factor at this bounce is equal to the unit, $a(t = 0) = 1$. 

Solving the effective Hamiltonian constraint in the mLQC-I model results in two asymmetric branches that represent the Universe evolution during the pre-bounce and post-bounce epochs. Each epoch is governed by its own modified equations, containing quantum corrections that are unique to this regularization. In particular, the bounce is characterized by a critical energy density that is given by 
$\rho_c^{I}=\rho_c/[4 (1+\gamma^2)]$.

In the mLQC-II model, the solution to the Hamiltonian constraint solution also involves two branches. However, only one of them is physically viable. This leads to a symmetric evolution with respect to the bounce, similar to what happens in the conventional LQC model, but governed by different evolution equations. The critical energy density is now greater than the conventional one. It is given by $\rho_c^{II}=4 (1+\gamma ^2) \rho_c $. Although the dynamics is governed by the corresponding modified cosmological equations, their analytical solutions for the scale factor are generally unavailable in closed form. Approximate solutions, similar to Eq. \eqref{eq:scalefactor-lqc}, have been proposed in Ref. \cite{Li:2019ipm}. For the purposes of our work, nevertheless, we prefer to compute the scale factor numerically in this model, restricting our approximations to the P\"oschl-Teller description of the background dependent mass of the perturbations, as we will explain in the next section.

In order to avoid apparent singularities in our numerical calculations for the mLQC-II model, we consider the second order linear differential equation
\begin{equation}\label{HII}
    \ddot{a}=\dot{H}a+H^2a, \qquad  H^2=\frac{8 \pi }{3}\rho  \left(1-\frac{\rho }{\rho_c^{II}}\right) \frac{  2+8 \left(1+\gamma ^2\right)
\gamma ^2 (\rho /\rho_c^{II})}{1+2 \gamma ^2 (\rho /\rho_c^{II})+\sqrt{1+4 \left(1+\gamma ^2\right) \gamma ^2 (\rho
/\rho_c^{II})}} ,
\end{equation}
with energy density that is given by $\rho(t)=\rho_c^{II}/a(t)^{6}$, similarly to what happens in conventional LQC. Our initial conditions at the bounce are $a(t=0)=1$ and  $\dot{a}(t=0)=0$. On the other hand, for the mLQC-I model, using that $\rho(t)a(t)^{6}$ also remains constant, we alternatively use the following second order equation for the energy density: 
\begin{equation}\label{eq:ddotrho}
\ddot{\rho}(t)=-6 \dot{H}\rho-6 H^2\rho,
\end{equation}
with initial conditions $\rho(t=0)=\rho_c^{I}$ and $\dot{\rho}(t=0)=0$. Then the scale factor is obtained as $a(t)=[\rho_c^{I}/\rho(t)]^{1/6}$. As we commented above, the evolution for this regularization is asymmetric with respect to the bounce. Thus, the Hubble parameter in Eq. \eqref{eq:ddotrho} for $t\leq 0$ corresponds to the pre-bounce branch, for which
\begin{equation}
    H^2_{pre}=\frac{1}{48 \pi ^2 \gamma ^2 \left(1+\gamma ^2\right)^2}\left(1-\frac{\rho }{\rho_c^{I}}\right) \left(1+\frac{(\rho  /\rho_c^{I}) \left(1-2 \gamma
   ^2+\sqrt{1-(\rho  /\rho_c^{I})}\right)}{4\gamma^2 \left(1+\sqrt{1-(\rho  /\rho
_c^{I})}\right)}\right),
\end{equation}
while for $t\geq 0$ it corresponds to the post-bounce branch,
\begin{equation}\label{HI}
    H^2_{post}=\frac{8\pi}{3}   \rho  \left(1-\frac{\rho }{\rho_c^{I}}\right) \left(1+\frac{\gamma ^2 (\rho / \rho_c^{I})}{\left(1+\gamma
   ^2\right) \left(1+\sqrt{1-(\rho /\rho_c^{I})}\right)^2}\right).
\end{equation}

\begin{figure}[h!]
\begin{minipage}[c]{.48\linewidth}
\centering
\includegraphics[width=1\linewidth]{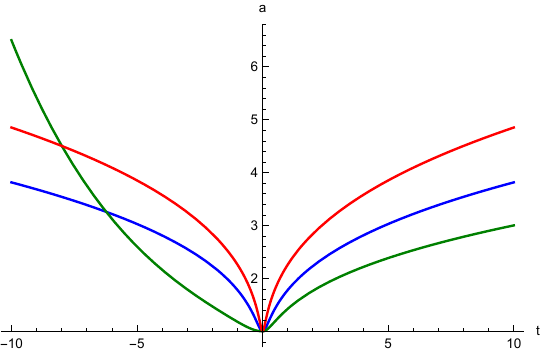}
\end{minipage}
\hfill
\begin{minipage}{.48\textwidth}
\centering
\includegraphics[width=1\linewidth]{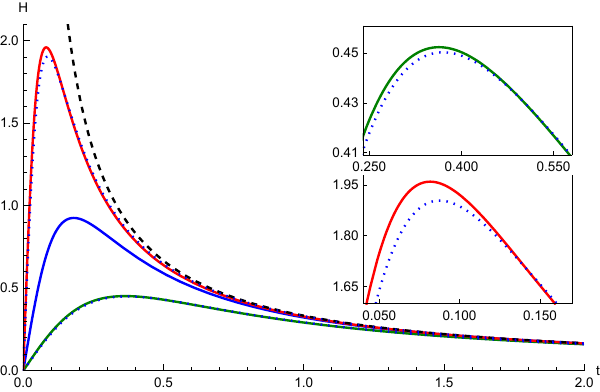}
\end{minipage}
\label{fig:qualitativebehaviour}
\caption{Scale factor $a$ (left panel) and Hubble parameter $H$ (right panel) in terms of the proper time $t$ for conventional LQC (blue solid line), the mLQC-I model (green solid line), and the mLQC-II model (red solid line). For comparison, the right panel also shows the classical evolution of $H$ for a kinetically dominated universe (black dashed line) and the two scaled Hubble parameters corresponding to conventional LQC (two blue dotted lines). To implement these scalings, $H^2$ has been multiplied by the critical density of the mLQC-I or mLQC-II model, respectively, and divided by the critical density of conventional LQC. Two insets are included around the maxima of the scaled Hubble parameters.}
\end{figure}

The solution for the scale factor with our initial conditions is shown in the left panel of Fig. \ref{fig:qualitativebehaviour} for the three considered regularizations. We see the difference in the cosmic evolution, depending on the regularization, as previously discussed. This difference is further illustrated in the right panel of the figure, where the corresponding Hubble parameters are compared, including that of a classical evolution in a matter-dominated universe ($H\approx 1/(3t)$). It is clear that, after some time $t_0$, the classical description of general relativity becomes valid for all three regularizations, although at different timescales in each case. Moreover, our calculations and plots indicate that the differences between the distinct quantum corrections arising in each of the regularizations are small. In fact, by scaling the time and the inverse of the Hubble parameter by the square root of the critical density of each of the three LQC models, the solutions turn out to match with great precision after the bounce. For example, for the Hubble parameter, the relative deviation after this scaling is less than $1\%$ according to our numerical calculations. Such scaling properties arise because the square Hubble parameter over the critical density differs in the three models (after the bounce) by a quantity of the order of $\gamma^2$ multiplied by the ratio of the energy density and the critical one, as can be seen from Eqs. \eqref{eq:Friedmann-LQC}, \eqref{HII}, and \eqref{HI}. Since the energy density ratio is always smaller than 1 and $\gamma^2$ is considerably small, the resulting Hubble equation after the commented scaling transformation is extremely similar. Hence, the scale factor obtained after the bounce from this equation and the conservation of $\rho a^{6}$ displays very accurately the noted scaling behavior.  

\section{Cosmological Perturbations}\label{cosmo}

We now turn our attention to the evolution of cosmological perturbations in the framework of LQC, analyzing their dynamics on the modified effective backgrounds described in the previous section. Several quantization schemes have been proposed for primordial perturbations in LQC, reflecting diverse methods for incorporating quantum geometry corrections into the early Universe dynamics. Among them, the dressed metric \cite{Agullo:2012sh, Agullo:2012fc,Agullo:2013ai} and the hybrid quantization approaches \cite{ElizagaNavascues:2020uyf,Gomar:2015oea} are particularly noteworthy owing to their ultraviolet behavior, similar to that of conventional inflationary models in general relativity.

In this work, we will focus our discussion on the hybrid quantization approach. In this framework, gauge-invariant (scalar) perturbations are governed by a quantum corrected Mukhanov-Sasaki equation \cite{Mukhanov:1988jd,Sasaki:1983kd}. These gauge-invariant perturbations can be represented by Fourier modes $v_k$, where $k$ is their (angular) wavenumber \cite{ElizagaNavascues:2020uyf}. They satisfy the dynamical equation 
\begin{equation}\label{eq:MS}
v_k''+(k^2+s)v_k=0, \qquad  s=-\frac{4\pi}{3}a(\rho-3P),
\end{equation}
where the prime denote differentiation with respect to a conformal time $\tau$ and $s$ is an effective time-dependent mass term encoding the background dynamics \cite{ElizagaNavascues:2020uyf}. Under the assumption that the energy density at the quantum epoch after the bounce is dominated by the kinetic contribution of the inflaton field, as it happens in the solutions of phenomenological interest that we are considering, this mass simplifies to \cite{Navascues:2021mxq}
\begin{equation}\label{eq:mass}
s_{kin}=\frac{8\pi}{3}\frac{\rho_c}{a^{4}}.
\end{equation}
However, even in this situation, one cannot solve analytically the mode equation \eqref{eq:MS} for the perturbations. To address this challenge, it has been proposed that the behavior of the background-dependent mass near the bounce can be modeled using a P\"oschl-Teller potential \cite{Wu:2018sbr}, a solvable potential in quantum mechanics known for its exact bound-state solutions. For  LQC, the proposal is then to work with an approximated mass term of the form \cite{Navascues:2021mxq}
\begin{equation}\label{eq:PT}
s_{PT}=\frac{U_0}{\cosh^{2}{[\alpha(\tau-\tau_b)]}}, \qquad \alpha=\frac{\text{arcosh}[(a_0)^2]}{\tau_0-\tau_b}, \qquad U_0 =\frac{8\pi}{3}\rho_c.
\end{equation}
The formulas for the parameters $\alpha$ and $U_0$ are obtained by imposing that the approximated and exact masses coincide at $\tau_b$, which marks the conformal time at the bounce, and at the transition time $\tau_0$ to the classical regime with kinetic domination. The constant $a_0$ is the scale factor at this transition time. Therefore, beyond $\tau_0$, and until the onset of inflation at a time $\tau_{inf}$, the mass adopts a classical relativistic evolution, according to the expression
\begin{equation}\label{eq:sGR}
s_{GR}=\frac{1}{4}\left(\tau -\tau_0-\frac{1}{2 H_0a_0} \right)^{-2}, \qquad H_0= \frac{\sqrt{U_0}}{a_0^{3}},
\end{equation} 
which ensures continuity of $s$ at the transition time. A correct choice of this time is important, because the accuracy of our approximations heavily depends on it. If it is chosen too small, the classical relativistic description is not sufficiently good around it. But if we choose a too large value for this time, the error committed with the P\"oschl-Teller approximation during the quantum LQC regime increases. To quantify the accuracy of the approximation for the time-dependent mass $s_{kin}$, we employ the relative difference function
\begin{equation}
\text{Err}=
\left\{
\begin{aligned}
     \frac{2 | s_{PT} - s_{kin} |}{s_{PT} + s_{kin}}&, \qquad \text{for} ~~\tau \in [ \tau_b, \tau_0 ], \\
    \frac{2 \lvert s_{GR} - s_{kin} \rvert}{s_{GR} + s_{kin}}&, \qquad  \text{for} ~~ \tau \in (\tau_0, \tau_{inf} ],
\end{aligned}
\right.
\end{equation}
with a threshold of  $4$\%  in the kinetically dominated relativistic region, while also keeping errors in the quantum region after the bounce as small as possible. 

\begin{figure}[h!]
\begin{minipage}[c]{.32\linewidth}
\centering
\includegraphics[width=1\linewidth]{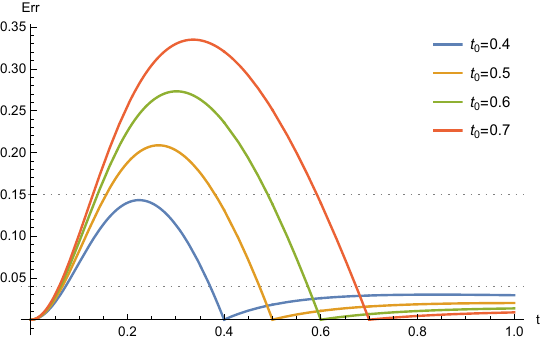}
\end{minipage}
\begin{minipage}[c]{.32\linewidth}
\centering
\includegraphics[width=1\linewidth]{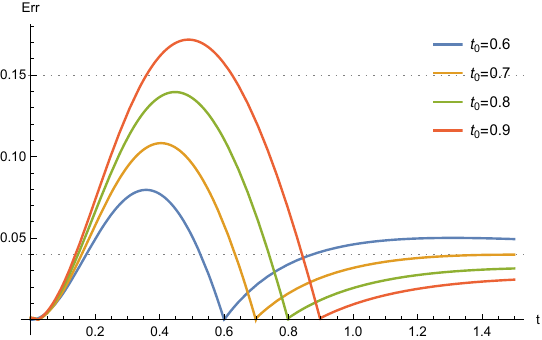}
\end{minipage}
\hfill
\begin{minipage}{.32\textwidth}
\centering
\includegraphics[width=1\linewidth]{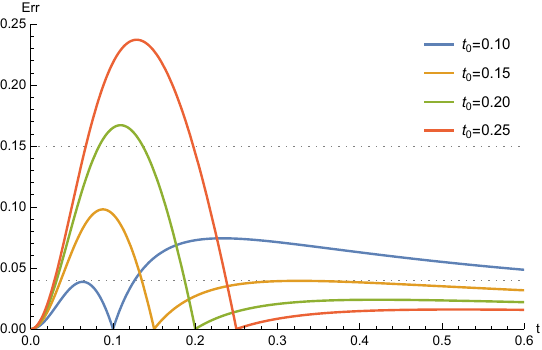}
\end{minipage}
\caption{Relative error $\text{Err}$ as a function of the proper time $t$. The left panel corresponds to the conventional LQC model, the middle panel to the mLQC-I model, and the right panel to the mLQC-II case.}
\label{fig:relativeerror}
\end{figure}

For conventional LQC, numerical optimization identifies $t_0 = 0.4$ (in Planck units) as the optimal transition proper time (corresponding to $\tau_0$), satisfying our requirements \cite{Navascues:2021mxq}. Our computations confirm that this time is an optimal choice, with a relative error that remains smaller than $15$\% in the entire quantum regime, as shown in Fig. \ref{fig:relativeerror}. We now want to check whether this P\"oschl-Teller approximation can as well be adopted for the other two regularizations. As we mentioned above, the effective background of the mLQC-I and mLQC-II models can be described remarkably well by accounting for the change of value of the critical density with respect to conventional LQC. As a result, we can deduce the optimal transition times for these models by an appropriate scaling of $t_0$ from conventional LQC with a factor given by the square root of the ratio between the respective critical densities of the considered models. This factor is $\sqrt{4 (1 + \gamma^2)}$ for mLQC-I, and its inverse for mLQC-II. Neglecting the small value of $\gamma^2$ compared to the unit, we then obtain the transition times $t_0^{I}= 0.8$ and $t_0^{II}= 0.2$ for mLQC-I and mLQC-II, respectively.  

Independently form the above estimate, by computing numerically the scale factors of the three studied models and the resulting background-dependent masses, we can calculate the relative error commited with our P\"oschl-Teller approximation with different choices of the transition time in each case. The results are displayed in Fig. \ref{fig:relativeerror}. The middle and right panels correspond to the mLQC-I and mLQC-II models, respectively.  In this way, we can confirm that our estimation of the optimal transition time is correct. 

Furthermore, we see that the behavior of the relative error for mLQC-I in the quantum period mirrors quite well that of conventional LQC. It might be a little bit surprising that the P\"oschl-Teller approximation for this model results to be so good, because it does not take into account the asymmetry of the background evolution with respect to the bounce. However, recall that we are setting our initial conditions at the bounce (the same for all the models) and the post-bounce history of the Universe is indeed very similar in mLQC-I and in conventional LQC, once one implements the scaling transformation discussed above.  

For mLQC-II, we can check that, with the suggested choice of transition time to the kinetically dominated relativistic region, the relative error in that region is kept below $4$\%. However, the relative error rises to a value close to $17$\% in the quantum region. This is not something unexpected, given the behavior of the Hubble parameter in Fig. \ref{fig:qualitativebehaviour}. The deviations in this parameter already indicate that the approximation obtained by simply scaling our physical quantities to account for the change of value of the critical density disregarding the value of $\gamma^2$ is not completely trustworthy in this case. However, the additional $3$\% of maximum relative error in the quantum epoch with respect to the conventional LQC case seems still reasonable, and does not drastically undermine the validity of our approximation. For this, and in order to make comparisons under scaling transformations easy to follow, we maintain our choice $t_0^{II}=0.2$ for the transition time of the mLQC-II model.

Before proceeding further, we want to make a brief comment about our approximation. By inspecting the mass term, as given in Eqs. \eqref{eq:PT} and \eqref{eq:sGR}, it is evident that it can be effectively scaled by $4 (1 + \gamma^2)$ or its inverse to obtain the results for the two alternative regularizations from conventional LQC, recalling the scaling properties of the background solutions that we have already noticed. Moreover, this scaling behavior affects as well the mode equations of the perturbations, given by Eq. \eqref{eq:MS}. Taking into account the commented transformations of the conformal time and the mass term, the mode equations remain invariant if the wavenumber $k$ is also scaled, in this case as the square root of the mass.  Thus, we can expect that the PPS of the mLQC-I and mLQC-II models display a similar power suppression as the spectrum of conventional LQC, but at a scaled wavenumber that would essentially correspond to the double or the half (ignoring again $\gamma^2$ compared to 1) of the suppression scale in conventional LQC. Since the general solution to the mode equations \eqref{eq:MS} with a P\"oschl-Teller potential as the effective mass has a known analytic expression \cite{Navascues:2021mxq}, we can obtain analytically the PPS if we can determine the solution associated with the proper choice of vacuum state.

At this point, it is worth recalling that the modes must be normalized using the Klein-Gordon inner product, which guarantees that the gauge-invariant variables satisfy the canonical commutation relations. This normalization amounts to imposing that
\begin{equation}
\mu_k(\mu_{k}^{'})^{*}- \mu_{k}^{'}\mu_k^{*}=i ,
\end{equation}
where the symbol $^*$ stands for complex conjugation. This condition is pivotal for constructing a vacuum state for the perturbations which provides the initial configuration and momentum for the modes near the bounce. In general, the behavior of the background in this effective quantum regime will drastically differ from a de Sitter expansion. Hence, a Bunch-Davies state is not a correctly justified choice of vacuum.\footnote{Although it is argued in the literature that it is possible to implement Bunch-Davies conditions in the asymptotic past of the pre-bounce branch in the mLQC-I model \cite{Li:2024xxz}, these conditions would evolve to different ones at the bounce. In addition, it is not clear whether imposing conditions on the pre-bounce branch is a good strategy, since this branch is extremely sensitive to the chosen regularization.} 

In the framework of LQC, a proposal that has been explored for an alternative choice of vacuum is to use the adiabatic approach \cite{Parker:1971pt}. This method, based on imposing adiabatic conditions on the perturbations, is particularly useful when exact solutions are not readily obtainable. The method breaks down in situations in which the mass term becomes negative and larger in absolute value than the square wavenumber of the perturbative mode (as it happens e.g. at the bounce for low wavenumbers in the dressed metric quantization). An alternative criterion for choosing a vacuum state in LQC involves minimizing a quantity related to the re-normalized stress-energy tensor \cite{Agullo:2014ica}. Nonetheless, the result depends strongly on the time when one requires the minimization. On the other hand, Ashtekar and Gupt have suggested to use the only state that has a Weyl curvature below a bound compatible with the uncertainty principle in the region with quantum effects, and displays a classical behavior at the end of inflation \cite{AGvacio2,AG}. 

In this work, we are going to follow instead a recently proposed criterion that is known to generalize the standard choice of the Poincar\'e vacuum for Minkowski spacetime and the Bunch-Davies state for de Sitter, permitting the extension to other more general backgrounds in cosmology \cite{ElizagaNavascues:2019itm}. In particular, this criterion has been applied successfully in the context of conventional LQC \cite{Navascues:2021mxq,ElizagaNavascues:2023xah,MenaMarugan:2024vyy,MenaMarugan:2024zcv}. It selects natural positive frequency solutions in the ultraviolet sector by demanding an asymptotic diagonalization of the Hamiltonian of the perturbations \cite{ElizagaNavascues:2019itm}. The privileged state is constructed from this set of solutions. More specifically, adopting for the perturbative modes an expression of the form
\begin{equation}
    \mu_k = \frac{1}{\sqrt{-2 \, \text{Im}(h_k)}} e^{i \int_{\tau_b}^{\tau} d\tilde{\tau} \, \text{Im}(h_k)(\tilde{\tau})} 
\end{equation}
with $h_k$ such that $\mu_k$ satisfies the mode equations, so that $h_k' = k^2 + s + h_k^2$, 
the diagonalization criterion chooses 
the solutions that admit the asymptotic behavior $k h_k^{-1} \sim  i + 2 i\sum [\gamma_n  /(2ik )^{n+2}] $, where $\gamma_n$ are time-dependent functions obtained by a recurrence formula from the value $\gamma_0=s$, once the mass term $s$ is given \cite{ElizagaNavascues:2019itm}. This criterion allows us to construct a physically meaningful state which is known to avoid spurious oscillations in the PPS
\cite{NO-Sant}. From now on we will refer to such a state as an NO-AHD vacuum (using the abbreviation explained in the Introduction). If we restrict our attention exclusively to the quantum epoch with P\"oschl-Teller mass, the solutions for $h_k$ are given by
\begin{equation}
   h_k= -i k+\frac{2 U_0}{\alpha +i k }x(1-x)\frac{\, _2F_1\left(\frac{3}{2}+\sqrt{\frac{1}{4} + \frac{U_0}{\alpha^2}},\frac{3}{2}-\sqrt{\frac{1}{4} + \frac{U_0}{\alpha^2}};2+\frac{i k}{\alpha
   };x\right)}{ \,
   _2F_1\left(\frac{1}{2} + \sqrt{\frac{1}{4} + \frac{U_0}{\alpha^2}},\frac{1}{2} - \sqrt{\frac{1}{4} + \frac{U_0}{\alpha^2}};1+\frac{i k}{\alpha };x\right)},
\end{equation}
which corresponds to the normalized mode solution
\begin{equation}\label{modePT}
    \mu^{PT}_{k} = \frac{1}{\sqrt{2k}} \left[ x(1 - x) \right]^{-ik/2\alpha} \, {}_2F_1 \left( \frac{1}{2} + \sqrt{\frac{1}{4} + \frac{U_0}{\alpha^2}}  - \frac{ik}{\alpha}, \frac{1}{2} - \sqrt{\frac{1}{4} + \frac{U_0}{\alpha^2}} - \frac{ik}{\alpha}; 1 - \frac{ik}{\alpha}; x \right),
\end{equation}
where ${}_2F_1$ is the hypergeometric function \cite{abramowitz1965handbook},  $x=1/[1 + e^{-2\alpha(\tau-\tau_b)}]$, and the constants $\alpha$ and $U_0$ are given in Eq. \eqref{eq:PT}. This would be the exact NO-AHD vacuum if the evolution of the Universe reduced to this quantum era around the bounce. However, the early Universe dynamics extend beyond it. Nonetheless, it can be argued (see e.g. Ref. \cite{Navascues:2021mxq}) that the main effect of the other preinflationary and inflationary regions is to introduce rapidly changing phases, which can be wisely removed by a suitable Bogoliubov transformation at the end of inflation. In the following we will use this strategy to determine a non-oscillating vacuum. 

At the transition time $\tau_0$, the classical description of a kinetically dominated period in general relativity becomes valid. The general mode solutions in this epoch are a linear combination of Hankel functions of the first and second kind \cite{abramowitz1965handbook,Navascues:2021mxq}, $H^{(1)}_0$ and $H^{(2)}_0$ respectively,
\begin{equation}\label{eq:KD}
    \mu_k^{GR} = C_k \sqrt{\frac{\pi y}{4}} H^{(1)}_0(ky) + D_k \sqrt{\frac{\pi y}{4}}H^{(2)}_0(ky), \qquad y = \tau - \tau_0 + \frac{1}{2 H_0 a_0}.
\end{equation}
The constants $C_k $ and $D_k$ appearing in this solution are determined by demanding the continuity of the mode solution and of its derivative at $\tau_0$, using Eq. \eqref{modePT} for $\tau<\tau_0$. One obtains
\begin{align}
C_k &= \left[H_0^{(1)}(\frac{k}{2k_0})\right]^{-1}
\left[
\sqrt{\frac{8k_0}{\pi}} \mu_k^{PT}(\tau_0) - D_k H_0^{(2)}\Big(\frac{k}{2k_0}\Big)
\right], \label{eq:Ck} \\
D_k &= i \sqrt{\frac{\pi}{8k_0}} 
\left[
k H_1^{(1)}\Big(\frac{k}{2k_0}\Big) \mu_k^{PT}(\tau_0) 
- k_0 H_0^{(1)}\Big(\frac{k}{2k_0}\Big) \mu_k^{PT}(\tau_0) + H_0^{(1)}\Big(\frac{k}{2k_0}\Big) (\mu_k^{PT})'(\tau_0)
\right], \label{eq:Dk}
\end{align}
where we have introduced a wavenumber scale $k_0=a_0H_0$, associated with the end of the effective quantum era after the bounce, and $ (\mu_k^{PT})(\tau_0)$ and $ (\mu_k^{PT})'(\tau_0)$ can be found with the expressions \cite{ElizagaNavascues:2020uyf}
\begin{equation}
 \mu_k^{PT}(\tau_0) = \sqrt{-\frac{1}{2\operatorname{Im}(h_k)(\tau_0)}}, \qquad 
 (\mu_k^{PT})'(\tau_0) = -h_k^*(\eta_0)\mu_k^{PT}(\tau_0).
\end{equation}

This kinetically dominated period is followed by a (slow-roll) inflationary period, that we approximate as a de Sitter expansion for simplicity (see Ref. \cite{ElizagaNavascues:2023xah} for the inclusion of slow-roll parameters). The mass term for this de Sitter background can be written as $s= - 2 k_{inf}^2 / \left[1- k_{inf} (\tau - \tau_{inf})\right]^{2} $, where we have introduced another wavenumber scale, $k_{inf}= H_{inf}a_{inf}$, related to the onset of inflation at $\tau=\tau_{inf}$. The general solution to the mode equations \eqref{eq:MS} in this case is \cite{Mukhanov:2005sc}
\begin{equation}\label{eq:sllowroll}
\mu_k=A_k \frac{e^{i k(\tau - {\bar{\tau}}_{inf})}}{\sqrt{2k}}\left[1 + \frac{i}{k(\tau  - {\bar{\tau}}_{inf})}\right] + B_k \frac{e^{-i k(\tau  - {\bar{\tau}}_{inf})}}{
\sqrt{2k}}\left[1 - \frac{i}{k(\tau - {\bar{\tau}}_{inf})}\right],
\end{equation}
where ${\bar{\tau}}_{inf}=\tau_{inf}-(1/k_{inf})$. As before, we demand continuity up to the first derivative in the mode solutions, now at time $\tau_{inf}$. Employing Eq. \eqref{eq:KD}, this fixes the constants $A_k $ and $B_k$ as
\begin{align}
\nonumber A_k &= 
\frac{e^{i k / k_{inf}}}{4\sqrt{k}} \sqrt{\frac{ \pi}{k_{inf}}}\Bigg\{ C_k \left[kH_0^{(1)}\left(\frac{k}{2 k_{inf}}\right) + \left(i k-k_{inf}\right) H_1^{(1)}\left(\frac{k}{2  k_{inf}}\right)\right] \\
&+ D_k \left[
k H_0^{(2)}\left(\frac{k}{2  k_{inf}}\right) 
+ \left(i k-k_{inf}\right)
H_1^{(2)}\left(\frac{k}{2 k_{inf}}\right)
\right]
\Bigg\}, \label{eq:Ak} \\
B_k &= A_k(ik \rightarrow -i k)  , \label{eq:Bk}
\end{align}
where the right hand side of the last line is the result of replacing $ik$ by $-ik$ in the terms where $k$ appears multiplied explicitly by the imaginary factor $i$ in Eq. \eqref{eq:Ak}. In addition, the constants $C_k $ and $D_k$ are given by Eqs. \eqref{eq:Ck} and \eqref{eq:Dk}, respectively. The derivation of these values of $A_k$ and $B_k$ follows the same procedure explained in Ref. \cite{Navascues:2021mxq}, particularly for the case of conventional LQC. It is therefore reasonable to anticipate that $A_k$ and $B_k$ exhibit a consistent ultraviolet behavior in the three considered models obtained with different regularizations. The variations among the three different cases are due to the different behavior of the cosmological background in the quantum period after the bounce. The specific peculiarities are encoded in the conformal time interval $\tau_0-\tau_b$ (or, equivalently, in the proper time $t_0$), in the two P\"oschl-Teller parameters $\alpha$ and $U_0$, and in the wavenumber scale $k_0=a_0 H_0$. We summarize their values in table \ref{tab:parameters}. We can check that, with a good degree of approximation, the values for the different models are really related by the scaling transformations that we have discussed: characteristic times scale similarly as the inverse square root of the critical density, $U_0$ as the critical density, and $\alpha$ and $k_0$ as its square root.   

\begin{table}[h!]
\caption{\label{tab:parameters}Values (in Planck units) of different parameters of the perturbations for the three regularizations}
\begin{ruledtabular}
\begin{tabular}{cccc}
\textbf{~~} & \textbf{conventional LQC} & \textbf{mLQC-I model} & \textbf{mLQC-II model} \\
\hline
\(t_0\) & \(0.40\) & \(0.80\) & \(0.20\) \\
\(\tau_0 - \tau_b\) & \(0.351\) & \(0.703\) & \(0.174\) \\
\(\alpha\) & \(3.421\)& \(1.797\) & \(7.122\)  \\
\(U_0\) & \(3.428\) & \(0.811\)& \(14.486\)  \\
\(k_0\) & \(0.932\)  & \(0.457\) & \(1.892\)\\
\end{tabular}
\end{ruledtabular}
\end{table}

\begin{figure}[h!]
\begin{minipage}[c]{.48\linewidth}
\centering
\includegraphics[width=1\linewidth]{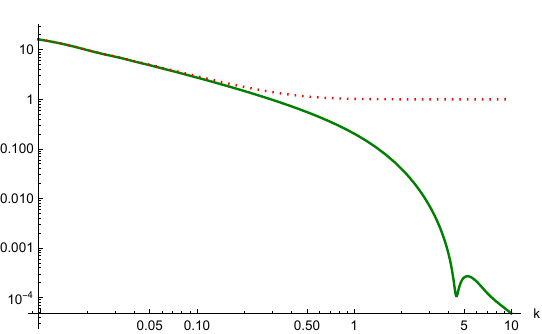}
\end{minipage}
\hfill
\begin{minipage}{.48\textwidth}
\centering
\includegraphics[width=1\linewidth]{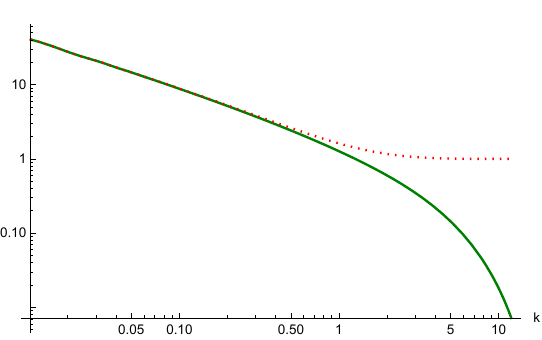}
\end{minipage}
\caption{Norms $|B_k|$ (dotted red line) and $|A_k|$ (solid green line) of the coefficients of the mode solutions. The left panel corresponds to the case of conventional LQC, while the right panel corresponds to the mLQC-II model. The two axes are in a logarithmic scale.}
\label{fig:norms}
\end{figure}

For effective solutions of phenomenological interest in LQC, the inflaton energy density at the onset of inflation is roughly of the order of $10^{-9}$ or a few orders of magnitude smaller (see e.g. Ref. \cite{ElizagaNavascues:2018bgp}). This provides an estimation of the wavenumber scale at the beginning of inflation.\footnote{The range of values with phenomenological interest leads to scales that are no more than one order of magnitude smaller.} For instance, for conventional LQC one can estimate that $k_{inf}=2.8\times 10^{-3}$.\footnote{Based on the scaling properties discussed above, one would then obtain $k_{inf}^{I}=1.4\times 10^{-3}$ and  $k_{inf}^{II}=5.6\times 10^{-3}$ for mLQC-I and mLQC-II respectively, all of them of similar order.} On the other hand, by examining the behavior of the coefficients $A_k$ and $B_k$ for conventional LQC, for the mLQC-I model, and for the mLQC-II model (the former and the latter displayed in Fig. \ref{fig:norms}), one can check that their complex norms approach the values $|A_k|\approx 0$ and $|B_k|\approx 1$ for large wavenumbers. Note that these limiting values satisfy the normalization condition $|B_k|^2=1+|A_k|^2$.

\section{Primordial Power Spectrum}\label{section-PPS}

\begin{figure}[h!]
\centering
\begin{minipage}{0.48\textwidth}
    \centering
    \includegraphics[width=\linewidth]{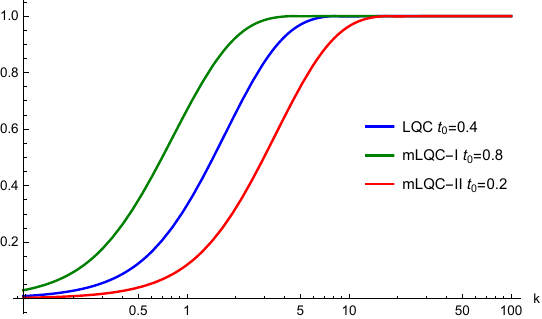}
\end{minipage}
\hfill
\centering
\begin{minipage}{0.48\textwidth}
    \centering
    \includegraphics[width=\linewidth]{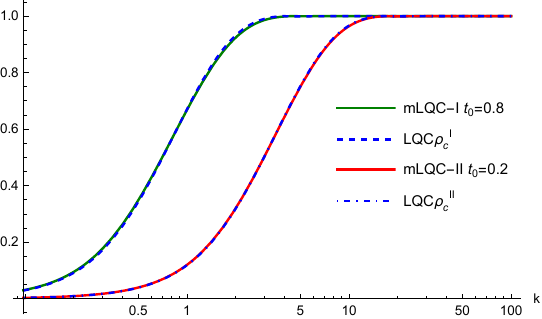}
\end{minipage}
\caption{Primordial power spectrum $4\pi^2\Tilde{\mathcal{P}}_{s}(k)/H_{inf}^2$ for the non-oscillating vacuum obtained from an asymptotic Hamiltonian diagonalization in the three considered models arising from different regularizations:  mLQC-I (green solid line), mLQC-II (red line), and conventional LQC (blue solid line in the left panel). In the right panel, we compare the spectra of the mLQC-I and mLQC-II models with the spectrum that would be obtained in conventional LQC if its critical density took the same value as in those models, $\rho^I_c$ and $\rho^{II}_c$ respectively (the blue dashed line corresponds to $\rho^I_c$ and the dot-dashed line to $\rho^{II}_c$).}
\label{fig:pps}
\end{figure}

We can now analytically compute the PPS of the perturbations for the three considered regularizations. In single-field inflationary cosmology, this PPS can be obtained from the evaluation of the perturbations at the end of inflation. For a de Sitter background, employing the general solution provided by Eq. \eqref{eq:sllowroll}, we see that the PPS can be expressed in terms of the coefficients $A_k$ and $B_k$ as 
\begin{equation}\label{pps}
\mathcal{P}_{s}(k)=\frac{H_{inf}^2}{4\pi^2}|B_k-A_k|^2 .
\end{equation}
The relative phase between $B_k$ and $A_k$ often introduces a rapid oscillation in $k$ that, upon averaging, leads to an enhancement of power. This oscillatory phase appears as a consequence of the loss of smoothness in our description of the mode solutions (which in general are only continuous up to the first derivative at the matching times between different dynamical periods). However, we can remove these phases and their power contribution by a convenient Bogoliubov transformation that restores a non-oscillating spectrum \cite{Navascues:2021mxq}. This Bogoliubov transformation simply sends $B_k$ and $A_k$ to their norms. Since these norms do not display rapid oscillations, neither does the resulting PPS. In this way, the PPS reached for a true NO-AHD state is
\begin{equation}\label{eq:nopps}
\Tilde{\mathcal{P}}_{s}(k)=\frac{H_{inf}^2}{4\pi^2}\left(|B_k|-|A_k|\right)^2 .
\end{equation}
Substituting the values given in Table \ref{tab:parameters} into Eqs.\eqref{eq:Ak}, and \eqref{eq:Bk} and using the above formula, we obtain the expression of the PPS for each of our three models. These spectra are displayed in Fig. \ref{fig:pps}. 

We can immediately identify distinct wavenumber scales corresponding to power suppression for the various regularizations  discussed here. These scales must be related by scaling transformations accounting for the change of critical density, according to our previous comments. Indeed, when we compute the corresponding scales, we see that they roughly coincide with $k_0$, which displays the aforementioned scaling behavior (see Table \ref{tab:parameters}). These scaling properties are confirmed by the right panel of Fig. \ref{fig:pps}, where we see that the PPS of the mLQC-I and mLQC-II models are almost identical to the PPS of conventional LQC with scaled critical energy density, with differences that are essentially ignorable for our purposes.

\begin{figure}[h!]
\centering
\begin{minipage}{0.48\textwidth}
    \centering
    \includegraphics[width=\linewidth]{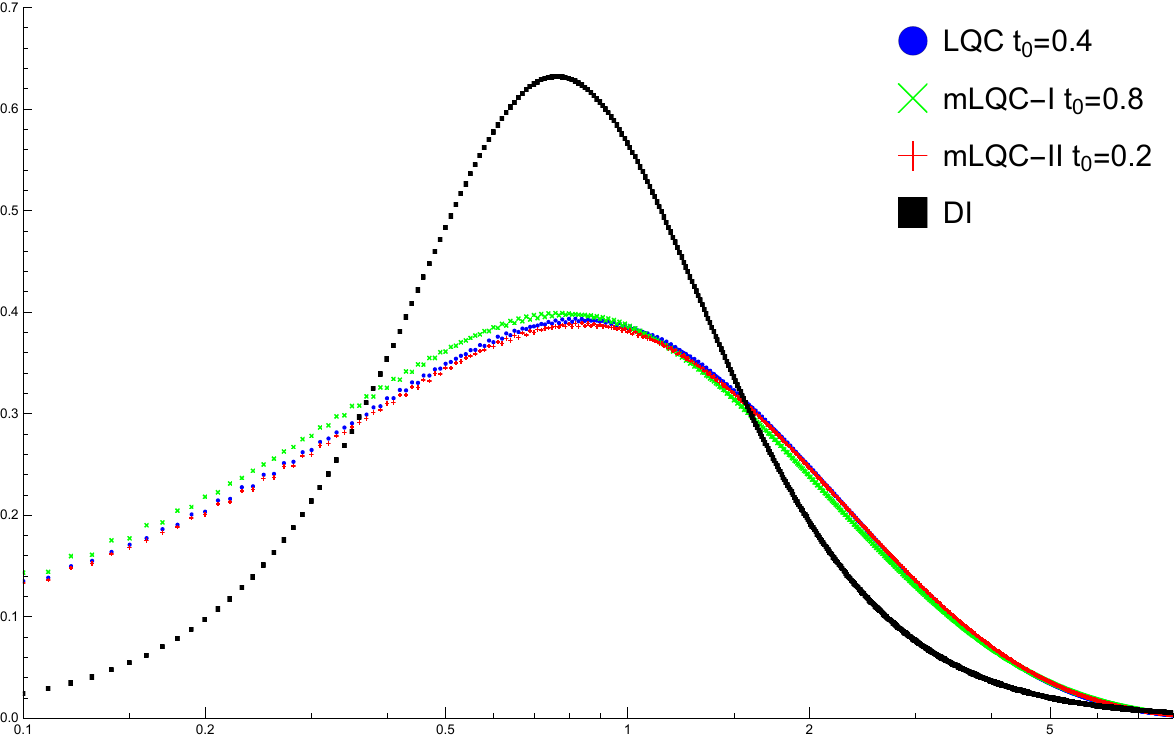}
\end{minipage}
\hfill
\centering
\begin{minipage}{0.48\textwidth}
    \centering
    \includegraphics[width=\linewidth]{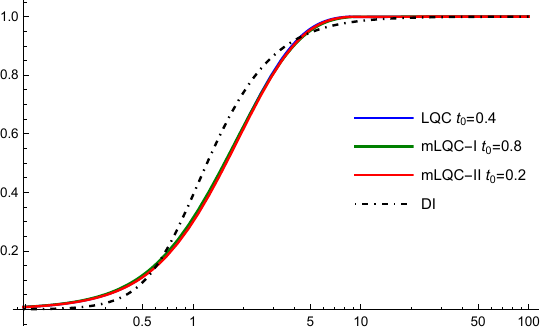}
\end{minipage}
    \caption{Left panel: First derivative of the primordial power spectrum $4\pi^2\Tilde{\mathcal{P}}_{s}(k)/H_{inf}^2$ for the different studied models. The derivative is taken with respect to the wavenumber divided by the square root of the critical density. Right panel: Primordial power spectrum $4\pi^2\Tilde{\mathcal{P}}_{s}(k)/H_{inf}^2$ in terms of the same scaled wavenumber as in the left panel. For comparison, we also show in both panels the case in which the initial quantum era corresponds to a de Sitter expansion with Planck energy density (black solid line, denoted as DI from the initials of delayed inflation).}
    \label{fig:slope}
\end{figure}

To further investigate the relationship between these spectra, we compare their slopes in the left panel of Fig. \ref{fig:slope}. Note that the scaled derivative of $4\pi^2\Tilde{\mathcal{P}}_{s}(k)/H_{inf}^2$ displayed in that panel is very similar for the three considered regularizations. While the general trend of power suppression remains consistent, this slope of the PPS is slightly greater for the mLQC-I model than for conventional LQC and the mLQC-II model. This tiny difference is a remnant of the slight discrepancy found in the background evolution during the quantum region after the bounce, even after performing a scaling transformation, and in our approximation of neglecting $\gamma^2$ in such a scaling. These two factors result in slight modifications of the parameters of the model with respect to those expected from the exact scaling behavior, in particular the parameters related to the transition to the classical regime to which the NO-ADH vacuum is relatively sensitive.  The right panel of Fig. \ref{fig:slope} confirms that the considered PPS (which we recall that has been normalized to the unit) is remarkably similar for the three regularizations when expressed in terms of a scaled wavenumber, equal to $k$ divided by the square root of the critical density. We also see in Fig. \ref{fig:slope} that the differences in the PPS (beyond the scaling behavior) are not ignorable with respect to the spectrum of a model with delayed inflation, in which the preinflationary era consists of a de Sitter expansion with Planck energy density followed by a kinetically dominated expansion. 

\section{Discussion}\label{section-Discussion}

We have analyzed the effects that different regularizations of the Hamiltonian constraint have on the PPS of the cosmological perturbations in the hybrid approach to LQC. In this analysis, the vacuum of the perturbations has been selected for all the regularizations by the NO-AHD criterion, that prescribes a state constructed from an asymptotic Hamiltonian diagonalization of the perturbations providing a non-oscillating spectrum. To model the background-dependent mass of the perturbations in the interval near (but after) the quantum bounce, we have utilized a P\"oschl-Teller approximation, which requires a careful determination of the transition time between the effective quantum regime and the classical relativistic era. To complete this step, we have had to compute the evolution of the scale factor for each of the discussed regularizations. In the conventional LQC model, the solution can be derived analytically. However, for the mLQC-I and mLQC-II models, which involve a more complicated background dynamics, we have used numerical methods to calculate the solution. With these solutions, we have estimated the value of the transition time to the classical behavior in each model, and then proceeded to resolve the corresponding Mukhanov-Sasaki equations for the perturbations, splitting the preinflationary and inflationary era in three intervals, namely, the quantum period after the bounce, a classical regime with kinetically dominated energy density of the inflaton, and a de Sitter (classical) regime. 

The resulting primordial spectra, presented in Sec. \ref{section-PPS}, display power suppression at low wavenumbers, and highlight the influence of the different regularizations in this suppression phenomenon. Notably, the scale of power suppression found for the mLQC-I and mLQC-II models can be accurately estimated by taking into account the change of critical density with respect to conventional LQC. In fact, our analysis reveals that this change approximately accounts for the differences among the PPS of the three models. Nonetheless, there are some small deviations from this approximate behavior. For instance, we have seen that the  slope of the PPS for the mLQC-I model is slightly steeper than what one would expect exclusively from a variation of the critical density, as illustrated in Fig. \ref{fig:slope}. However, the aforementioned deviations seem to be too tiny to be measurable with CMB observations within our range. From this perspective, the main parameter distinguishing the power suppression experimented with each of the three distinct regularizations is the value of the critical density, which is directly related with the wavenumber scale where this suppression ends for the NO-AHD vacuum. 

To conclude our discussion, it is illuminating to compare our results with those obtained in a related scenario in general relativity that incorporates the idea of a delayed inflation. This idea suggests that the period of slow-roll inflation may have been delayed owing to changes in the dynamics of the early Universe, leading to modifications of the traditional inflationary framework. Such alterations can leave unique signatures in the PPS, including the appearance of a cutoff scale. In particular, one may consider a model in which the Universe initially undergoes a (quasi) de Sitter phase during the Planck regime, followed by a transition epoch where the kinetic energy of the inflaton field dominates, and eventually leading to the standard slow-roll inflationary phase. In this model, it is possible to set initial conditions for the modes of the gauge-invariant perturbations in the Planck region by using again the NO-ADH criterion. These conditions select a Bunch-Davies state during the initial de Sitter expansion with Plank energy density. The resulting formulas for the coefficients $A_k$ and $B_k$ are similar to those in Eqs. \eqref{eq:Ak} and Eq. \eqref{eq:Bk}, but with different values of $C_k$ and $D_k$. The values for this model with delayed inflation can be found in Ref. \cite{PhysRevD.110.103502}. This change, arising from the distinct background dynamics during the Planck period and the corresponding modification of the privileged state that plays the role of a vacuum, leads to power suppression at a wavenumber scale which is roughly 40 times greater than in conventional LQC. When plotting the normalized PPS, $4\pi^2\Tilde{\mathcal{P}}_{s}(k)/H_{inf}^2$, to compare it with the PPS of the three LQC models (see right panel of Fig. \ref{fig:slope}), it becomes clear that the delayed inflationary model exhibits a different behavior. As shown in the left panel of Fig. \ref{fig:slope}, the slope of its normalized PPS (with respect to the wavenumber divided by the corresponding suppression scale, for which the slope reaches a maximum) is markedly steeper than that of the LQC models, proving the importance of the specific quantum corrections, which are different in this model with delayed inflation and in LQC.

\acknowledgments

The authors are grateful to B. Elizaga Navascu\'es for discussions. This work was supported by the Spanish grant PID2023-149018NB-C41 funded by MCIN/AEI/10.13039/501100011033/ and by the Polish National Center for Science (Narodowe Centrum Nauki -- NCN) under grant OPUS 2020/37/B/ST2/03604.

\end{document}